\documentclass[aps,prb,twocolumn,showpacs,superscriptaddress,floatfix]{revtex4}

\usepackage{graphicx,rotating}

\newcommand{\surf}[2]{(#1$\times$#2)\relax}

\begin{document}

\title{Exact location of dopants below the Si(001):H surface from scanning tunnelling
microscopy and density functional theory}

\author{Veronika Br\'azdov\'a}
\email[]{v.brazdova@ucl.ac.uk}
\thanks{Corresponding author}
\affiliation{London Centre for Nanotechnology, UCL, 17--19 Gordon St, London, WC1H 0AH, U.K.}
\affiliation{Department of Physics \& Astronomy, UCL, Gower St, London, WC1E 6BT, U.K.}
\affiliation{TYC@UCL, Gower St, London, WC1E 6BT, U.K.}

\author{David R. Bowler}
\email[]{david.bowler@ucl.ac.uk}
\affiliation{London Centre for Nanotechnology, UCL, 17--19 Gordon St, London, WC1H 0AH, U.K.}
\affiliation{Department of Physics \& Astronomy, UCL, Gower St, London, WC1E 6BT, U.K.}
\affiliation{TYC@UCL, Gower St, London, WC1E 6BT, U.K.}

\author{Kitiphat Sinthiptharakoon}
\affiliation{London Centre for Nanotechnology, UCL, 17--19 Gordon St, London, WC1H 0AH, U.K.}
\affiliation{Department of Electronic and Electrical Engineering, UCL, Torrington Place, London, WC1E 7JE, U.K.}

\author{Philipp Studer}
\affiliation{London Centre for Nanotechnology, UCL, 17--19 Gordon St, London, WC1H 0AH, U.K.}
\affiliation{Department of Electronic and Electrical Engineering, UCL, Torrington Place, London, WC1E 7JE, U.K.}

\author{Adam Rahnejat}
\affiliation{London Centre for Nanotechnology, UCL, 17--19 Gordon St, London, WC1H 0AH, U.K.}
\affiliation{Department of Physics \& Astronomy, UCL, Gower St, London, WC1E 6BT, U.K.}

\author{Neil J. Curson}
\affiliation{London Centre for Nanotechnology, UCL, 17--19 Gordon St, London, WC1H 0AH, U.K.}
\affiliation{Department of Electronic and Electrical Engineering, UCL, Torrington Place, London, WC1E 7JE, U.K.}

\author{Steven R. Schofield}
\affiliation{London Centre for Nanotechnology, UCL, 17--19 Gordon St, London, WC1H 0AH, U.K.}
\affiliation{Department of Physics \& Astronomy, UCL, Gower St, London, WC1E 6BT, U.K.}

\author{Andrew J. Fisher}
\affiliation{London Centre for Nanotechnology, UCL, 17--19 Gordon St, London, WC1H 0AH, U.K.}
\affiliation{Department of Physics \& Astronomy, UCL, Gower St, London, WC1E 6BT, U.K.}
\affiliation{TYC@UCL, Gower St, London, WC1E 6BT, U.K.}

\date{\today}

\begin{abstract} 
Control of dopants in silicon remains crucial to tailoring the properties of electronic materials for integrated circuits.   Silicon is also finding new applications in coherent quantum devices, as a magnetically quiet environment for impurity orbitals. 
The ionization energies and shapes of the dopant orbitals depend on the surfaces and interfaces with which they interact. The location of the dopant and local environment effects will therefore determine the functionality of both future quantum information processors and next-generation semiconductor devices.
Here we match observed dopant wavefunctions from scanning tunnelling microscopy (STM) to
images simulated from first-principles density functional theory (DFT) calculations, and precisely determine the substitutional sites of neutral As dopants between 5 and 15~\AA\/ below the Si(001):H surface.
We gain a full understanding of the interaction of the donor state with the surface, and the transition between the bulk dopant  and the dopants in the surface layer.
\end{abstract}

\maketitle

\section{Introduction\label{sec:introduction}}
A key development in the ongoing transformation of the doping of silicon from a random process to a deterministic one has been the discovery that individual phosphorus atoms can be positioned on silicon surfaces with atomic precision by means of scanning tunnelling microscopy (STM)\cite{a:tucker1998,a:Schofield2003,a:Oberbeck2004} and buried beneath controllable depths of silicon. 
This has opened the way to the fabrication of atomically patterned dopant devices beyond simple $\delta$-doped layers, \cite{a:Oberbeck2002,a:Fuhrer2006} to quantum atomic-scale nanowires,\cite{a:Weber2012,a:ruess2008,a:mckibbin2013} quantum dots,\cite{a:Fuechsle2010} and a single-dopant single-electron transistor,\cite{a:Fuechsle2012} all of which are formed from buried dopants within $\sim$25~nm of the surface. It also holds potential for the realisation of theoretical proposals to process quantum information encoded in the impurity spin states using either electrical \cite{a:Kane1998} or optical \cite{a:Stoneham2003,a:greenland2008} control of the donor wavefunctions.  The recently demonstrated readout \cite{a:morello2010} and control \cite{a:pla2013} of the electron and nuclear spin states of single P donors in Si highlight the enormous potential benefits of deterministic doping at the atomic-scale.

For effective coupling to gates for spin initialisation, readout, or electrical control, the dopants must be positioned close to a growth surface; similarly, proximity to surfaces or interfaces is an important factor in sub-10~nm CMOS devices. In such locations dopants are intermediate between two well-understood limits. For bulk donors~\cite{a:Brazdova2015} a diffuse hydrogenic state well-described by effective-mass theory \cite{a:luttinger1955,a:koiller2004} carries an unpaired electron spin in the neutral state, while for donors incorporated in the surface layer, the properties are strongly influenced by the local chemistry of the surface. \cite{a:radny2006,a:radny2007,a:Studer2012} 

Here we investigate arsenic donors in this critical region where the donors retain their essential hydrogenic character but are strongly perturbed by their proximity to a vacuum interface. In particular, we map the projected local density of states of As donors up to 12 atomic layers beneath a hydrogen-terminated silicon (001) surface using scanning tunnelling microscopy and spectroscopy (STM/STS).  We observe highly anisotropic features superimposed on the surface atomic lattice that we have interpreted as arising from the ground state wave function of neutral donors in analogy with previous reports of dopant wave function mapping in GaAs.\cite{a:Yakunin2004,a:Jancu2008} For donor depths beyond 12 layers we find even more diffuse surface features in agreement with a recent report where such features were assigned to As donors approximately 20 atomic layers beneath the surface based on bulk $\mathbf{k}\cdot\mathbf{p}$ and tight-binding calculations\cite{Salfi2014,Salfi2015}; more recent reports\cite{Usman2016,Saraiva2016} show that both these approaches give excellent agreement with experiment.  This type of semi-empirical approach allows for large-scale calculations and will correctly capture the long-range behaviour of the dopant wavefunction, which is particularly important for dopants far from the surface.  However, unlike these previously published reports, we employ a fully self-consistent, \emph{ab inito} computational method, which provides a much more detailed picture of the wave function-surface interactions and their importance.

\begin{figure*}
\includegraphics[width=16.0cm]{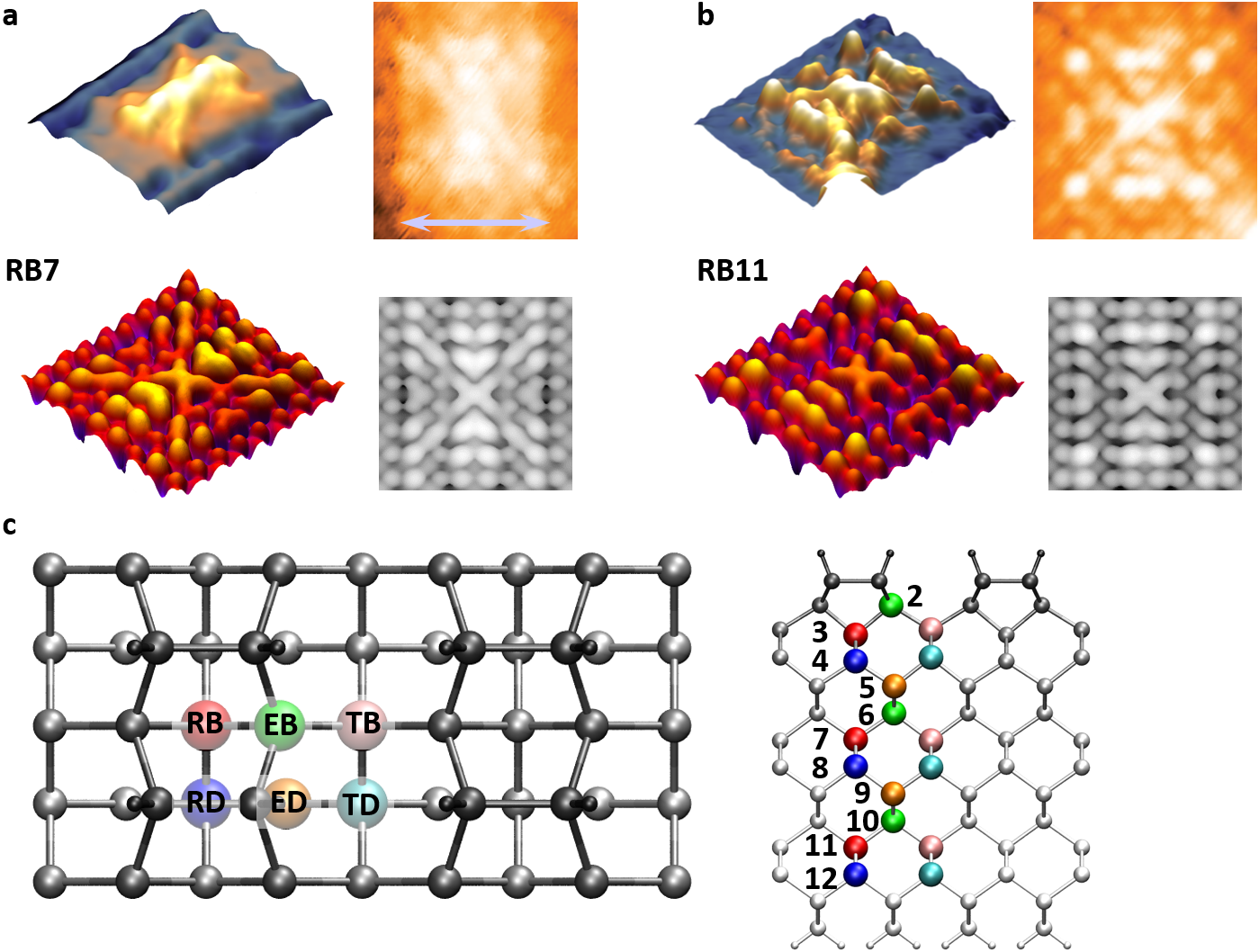}%
\caption{\label{f:fig1}(Color online) (a) and (b) Experimental (top) and simulated (bottom) STM images of occupied states of two 
dopants with the same surface unit cell symmetry (RB) but at different depths (layers 7 and 11, 
respectively). The area shown for the simulated images is 30.7~\AA$\times$30.7~\AA.
The arrow in the experimental image marks three dimer rows, $\sim$23~\AA. For both features the same image is rendered in 3D (left) and 2D (right); in 3D images the dimer rows run bottom left to top right, while in 2D images they run top to bottom. 
Dopant depth: (a) 9.5~\AA, 
(b) 15.0~\AA. The experimental images were recorded at sample bias voltages of $-$1.2~V (a) and $-$1.3~V (b). The simulated images were integrated from $E_F + 0.086$~eV to $E_F - 0.7$~eV.
(c) The slab model and site labelling used in this work. Left: symmetry labels for the six sites. Right: layer numbers.} 
\end{figure*}

Furthermore, we build on these results: by combining STM imaging with DFT, we are able to determine the precise lattice site of individual As dopants between 3 and 12 layers below the H:Si(001) surface (0.41 and 1.36~nm beneath the surface hydrogens, respectively). This is possible because our first-principles calculations explicitly describe the surface electronic structure, enabling us to establish directly the precise registry and depth of the defects through their interaction with the surface structure.  We find that the defects we study are relatively close to the surface and that the interaction of their wavefunctions with the surface reconstruction causes very specific symmetry breaking that gives the STM images their characteristic appearance.  We illustrate this in Fig.~\ref{f:fig1}, which shows experimental and simulated images of two different sites with the same surface symmetry: we assign the dopants in Fig.~\ref{f:fig1}a and Fig.~\ref{f:fig1}b to sites 7 and 11 layers below the surface H atoms, respectively.   The difference in depth is just 5.4~\AA, yet we can clearly distinguish the two different sites.  Because our calculations give a full description of the electronic state associated with each site,  we can also understand the evolution of the electronic structure from the bulk to the surface limit in detail and locate individual dopants on that pathway.  

In our previous work \cite{a:Kitiphat2014} we identified a set of asymmetric features that we attributed to subsurface As dopants and determined they were in the neutral charge state, similar to deeper positioned dopants observed under certain imaging conditions.\cite{Salfi2014,a:Voisin2015} In total, 122 neutral dopants were observed in the 285 images (each 50 nm $\times$ 50 nm) of the surface that we obtained, on several different days. In this work, we use DFT to identify the precise location of the dopant atoms with respect both to their depth below the surface and their registry with the surface reconstruction, and we show that within the top 12 layers the dopants interact significantly with the surface.  
Our interpretation of the image data is that the dopants are either
in the neutral or negative charge-state in filled-state images
(negative sample bias), allowing their wavefunctions to be mapped,
noting that if the donors are negatively charged then the imaging
contrast from the dopant dominates over the screened coloumb potential.
Examination of figure 2 in Ref.\onlinecite{a:Kitiphat2014} confirms there is not a strong
influence of any screened Coulomb potential. When imaged in empty
states (positive sample bias), the dopants become positively ionized,
owing to the influence of tip-induced band bending.
In the filled states the As dopants exhibit very distinctive image features that are large compared to the interatomic lattice spacing.  Analysis of the symmetry of each of the features allows identification of the positions of the dopant atoms within the surface unit cell, while subtle differences between features of the same symmetry yields information regarding their depth beneath the surface.  
We simulated STM images for all sites within the first twelve layers and
we show that we can give a complete account of the electronic and structural properties of these dopants.

\section{\label{s:methods}Methods}
We used an Omicron LT-STM system and measurements were performed at 77~K with a base pressure below $5\times10^{-11}$~mbar.  
STM tips were prepared by electrochemically etching 0.25 mm diameter polycrystalline W wire before loading into the vacuum system, followed by in-vacuum electron-beam annealing and field emission. Tips demonstrating reproducible field emission characteristics typically performed reliable, reproducible STM spectroscopy.
The STM images were obtained using multiple tips, which can give 
variability in both the voltage and the contrast relating to dopant
appearence; only selected images are used to identify dopants.
We used commercial Si(001) wafers (Compart Technology Ltd), highly doped with As with room temperature resistivity of 0.0015--0.004~$\Omega$~cm. The samples were first degassed for 12~h in UHV at $\sim$500~$^\circ$C, \textbf{}then flash annealed at $\sim$1050~$^\circ$C for 10~s and finally cooled down from 800~$^\circ$C to room temperature within 200~s. We avoided repeated sample flashes which
can lead to preferential desorption of n-type dopant atoms from the Si(001) surface region.\cite{a:Wolkow2012} Clean Si(001) surfaces were exposed to a beam of atomic hydrogen (Tectra atomic hydrogen source) for 5~min at a substrate temperature of $\sim$400$^\circ$C to produce monohydride terminated surfaces. All images were obtained with a set point current of 20~pA and are displayed with the same intensity (height) scale. STM images were processed with a combination of plane and line subtraction algorithms, and corrected for distortions (skew and creep).  

Density functional theory with the gradient-corrected Perdew-Burke-Ernzenhof 
(PBE)\cite{a:PBE} exchange-correlation functional, as implemented in the VASP\cite{a:VASP1,a:VASP2} code, was used. The core 
electrons were described by the projector augmented-wave (PAW) 
method.\cite{a:PAW1,a:PAW2}
The plane-wave basis set kinetic energy cutoff 
was set to 200 eV. The Brillouin zone was sampled using a (2$\times$2$\times$1) Monkhorst-Pack 
grid.\cite{a:Monkhorst1} Gaussian smearing was used for fractional
occupancies, with an 0.1\,eV width.
The convergence criterion for forces on atoms was 0.01~eV/\AA\/ and for total energy 10$^{-6}$ eV. 

We used the experimental bulk lattice parameter ($a$=5.431~\AA),\cite{a:Hubbard1975,a:Massa2009}
which is only trivially different (0.7\% larger, Ref.~\onlinecite{a:Brazdova2010b}) from the calculated PBE constant.
The surface unit cell was chosen so that the Si dimers on the surface were parallel
to one of the cell vectors, with a surface lattice parameter of $a/\sqrt{2}$=3.86~\AA.
We used a \surf88 surface unit cell, resulting in eight dimers in each dimer
row and four dimer rows in one computational cell. The lattice parameter perpendicular to the surface was set to 32~\AA, which results in approximately 12~\AA\/ of vacuum between the 
periodic images of the slab. The slab contained 14~Si atomic layers and was 
terminated with hydrogen atoms at the top (monohydride) and bottom (dihydride) surfaces, giving 1088 atoms in total.
The bottom two Si layers were kept fixed at bulk positions, the rest of the structure
was optimised for each system. 

In the experimental samples the resistivity at 300\,K is between 
$1.5\times 10^{-5}\,\Omega\,\mathrm{m}$ and $4.0\times 10^{-5}\,\Omega\,\mathrm{m}$, corresponding to a bulk 
dopant density\cite{Masetti1983} $n\approx10^{19}\,\mathrm{cm}^{-3}$. Although well above the metal-insulator transition, this is nevertheless such that the (randomly sited) donors are relatively well isolated (in the surface, all donors imaged were at least 3nm from any other donor\cite{a:Kitiphat2014}).  In our calculations the periodic boundary conditions mean that there is an ordered square array of one donor per surface supercell; taking the 
effective slab thickness as 14 layer spacings ($19.0\,\mathrm{\AA}$) gives an effective doping density of $n_\mathrm{eff}\approx 5.57\times 10^{18}$~cm$^{-3}$.   Consequently the donor states in 
our calculations carry a well-defined crystal momentum index $\mathbf{k}_\parallel$ parallel to the surface and are somewhat broader than those in the experiments; details of band-structure are given in Appendix~\ref{sec:bandstructures}. In an \emph{n}-type sample under filled-states imaging conditions (negative sample bias) the bands will be bent down near the surface so that some conduction-band states become occupied,\cite{a:brown2002,a:radny2006} populated by electrons thermally excited from the large ensemble of bulk donors. The resulting negative surface charge density screens the electric field from the tip. In our calculations we simulate this effect by adding one additional electron per unit cell to the system, corresponding to an areal charge density 
$\sigma=-1.06|e|\times 10^{17}\,\mathrm{m}^{-2}=-1.70\times 10^{-2}\,\mathrm{C}\,\mathrm{m}^{-2}$.  In the 
experiments this charge would screen a local electric field $\mathcal{E}=\frac{|\sigma|}{\epsilon_0}\approx 1.92\times 10^9\,\mathrm{V}\,\mathrm{m}^{-1}$ 
perpendicular to the surface, which is of the correct order of magnitude for a tip-sample bias of around 1\,V at a tip-surface spacing of around 5~\AA; we note that, as is standard, neither the tip nor the field is included, and this estimate confirms that our approach is consistent with the experimental conditions.  In our calculations, the negative surface charge is compensated by a uniform positive charge density over the entire simulation cell, giving overall neutrality as required by periodic boundary conditions.  The (unphysical) interaction between the positive and negative charge densities is the same for all the calculations we report and can therefore be neglected in the calculation of energy differences, geometries, and electronic states.

STM images were computed from the local densities of states
using the Tersoff-Hamann approach.\cite{a:tersoff-hamann1985}
We find the best agreement with experiment when we include the partially occupied dopant bands within the fractional-occupancy smearing width above the Fermi level, as well as a window of the fully occupied states. 
Experimental filled-state images are obtained with negative sample bias; hence in simulated images, electrons should tunnel from a window of states just below the Fermi energy (from $E_F$ to $E_F-\mathrm{e}V_\textrm{bias}$ at low temperature, where $E_F$ is the computational Fermi energy).  The $(2\times 2\times 1)$ Monkhorst-Pack $k$-point grid used elsewhere in this paper does not include the $\Gamma$-point and the lowest states therefore lie just above $E_F$; hence an STM integration window with an upper limit set to the Fermi energy $E_F$ does not contain any bands at the sampled $k$-points and does not produce a useful STM image. We have therefore shifted the integration window, typically so its upper bound is at $E_F+0.086~\mathrm{eV}$.
It is expected that the occupancies of states near the STM tip are driven some way from equilibrium by the complex balance of electron transfer rates to the bulk and to the tip and it is therefore not surprising that they contribute differently to the STM image.  Full details of how the Fermi level and k-point sampling settings affect the simulated images are given in Appendix~\ref{sec:effect-fermi-level}.

3D images were rendered with Gwyddion~2.36 (Ref.~\onlinecite{a:Necas2012}).

\section{Results and Discussion\label{sec:results-discussion}}
The 2$\times$1 surface unit cell of the hydrogen-terminated Si(001) surface provides six possible substitutional subsurface sites, distinguished by the relative position of the dopant with respect to the dimer rows on the surface and to the dimers themselves (Fig.~\ref{f:fig1}c).
With respect to the dimer rows, an atom can be either in the middle 
of a dimer row (R), in the middle of a trench between two dimer rows (T), or at the edge of 
a dimer row and a trench (E). In any of these locations an atom can be positioned in line 
with a dimer (D) or in between two dimers (B). 
Each of these six sites repeats every four layers (i.e., once per bulk unit cell) giving rise to a \textit{family} of possible structures having the same symmetry with respect to the surface.  The possible combinations of these locations give
the six symmetry families (see Fig.~\ref{f:fig1}(c)) . We number the 
atomic layers from 1 to 12 starting with the Si atoms of the dimers, and append the number to 
the site family. For instance, TD4 denotes a dopant atom located in the middle of a \textit{trench} 
between two \textit{dimer rows}, in line with a dimer, in \textit{layer 4}. Thus, layers 3, 7, and 11 are symmetry-equivalent in the surface unit cell,
as are layers 4, 8, and 12, and layers 5 and 9.   

\begin{figure*}
\includegraphics[width=18.0cm]{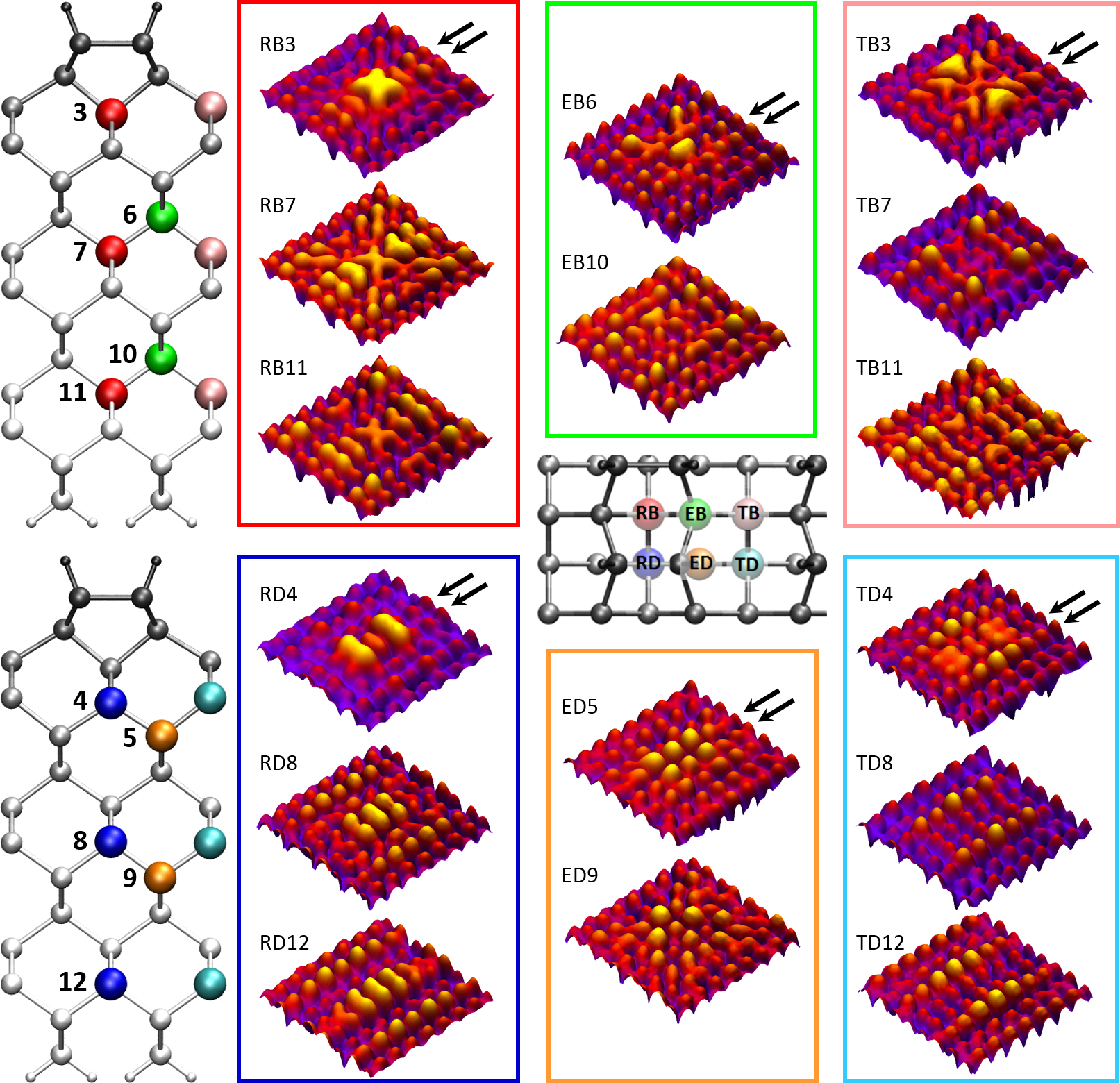}%
\caption{\label{f:fig2}(Color online) Simulated STM images of occupied
states for each dopant site, grouped by site symmetry (the colour code for each symmetry pane matches the corresponding colour of the sites on the structure on the left). 
The pairs of black arrows denote a dimer row. For details of the integration window of the simulated images see Appendix~\ref{sec:effect-fermi-level}.
}
\end{figure*}

In Fig.~\ref{f:fig2} we show the simulated STM images for every dopant position: each
symmetry family shares common motifs, though the relative weights of the motifs
vary with depth.  The two in-row positions (RD and RB) both show bright
protrusions symmetric about the dimer row, though with very different symmetry along the row.  The two 
trench positions (TD and TB) both show bright dots in neighbouring dimer rows,
but differ in the number of dots.  The edge positions (ED and EB) have mirror symmetry only in a plane perpendicular to the dimer
rows.  Notice that dopants within one symmetry family have significant differences that will allow us to assign the exact
atomic layer to the different dopants.  

In order to identify the features observed in STM, we compare them with DFT-simulated STM images.  To this end we now present a brief description of the appearance of each of the six families of structures; these are the criteria by which we assign experimental images to a particular dopant site.  

The RB sites share a distinctive $\times$-shape in the centre of the features,
centered between dimers and on the dimer row. The $\times$ becomes fainter with
increasing dopant depth, due to the different parts of the wavefunction
interacting with the surface. The central dimer row is slightly brighter than
the neighbouring rows; smaller features radiate from the $\times$ along
the central row and diagonally to the adjoining rows: the general effect is of a
larger $\times$-shape.

The RD sites all have three central bright features coincident with three
adjacent dimers in the same dimer row. There are five additional, smaller
features on each of the neighbouring dimer rows. These outer features become brighter
relative to the three central features as the dopant depth increases.

The EB and ED features are not symmetric about the dimer row, but instead have a v-shaped appearance. The main
distinguishing characteristic between the two families is the position of the
centre of the feature, as expected from the location of the dopant: EB sites are
centered between two dimers of the same row, ED sites on a dimer. 

The TB and TD sites are centered between dimer rows. They present small,
bright features along the two adjacent dimer rows, appearing as two parallel
rows of an even (TB) and odd (TD) number of bright spots. The TB sites have two
small v-shaped features at either side of the dimer trough which become significantly fainter as the
dopant depth increases: they are essentially the $\times$-shape of the RB sites,
split between two dimer rows. The TD sites do not have a strong central feature.
Within the TD family the sites are distinguished by the relative intensities of
the bright dots: the central dot is brightest in all; in TD4 and TD12 the two
subsequent dots (on both sides) are next in brightness, while in TB8 the
brightness alternates along the dimer row. The TD4 site differs from TD12 in
that TD4 has a brighter centre.
\begin{figure*}
\includegraphics[width=15cm]{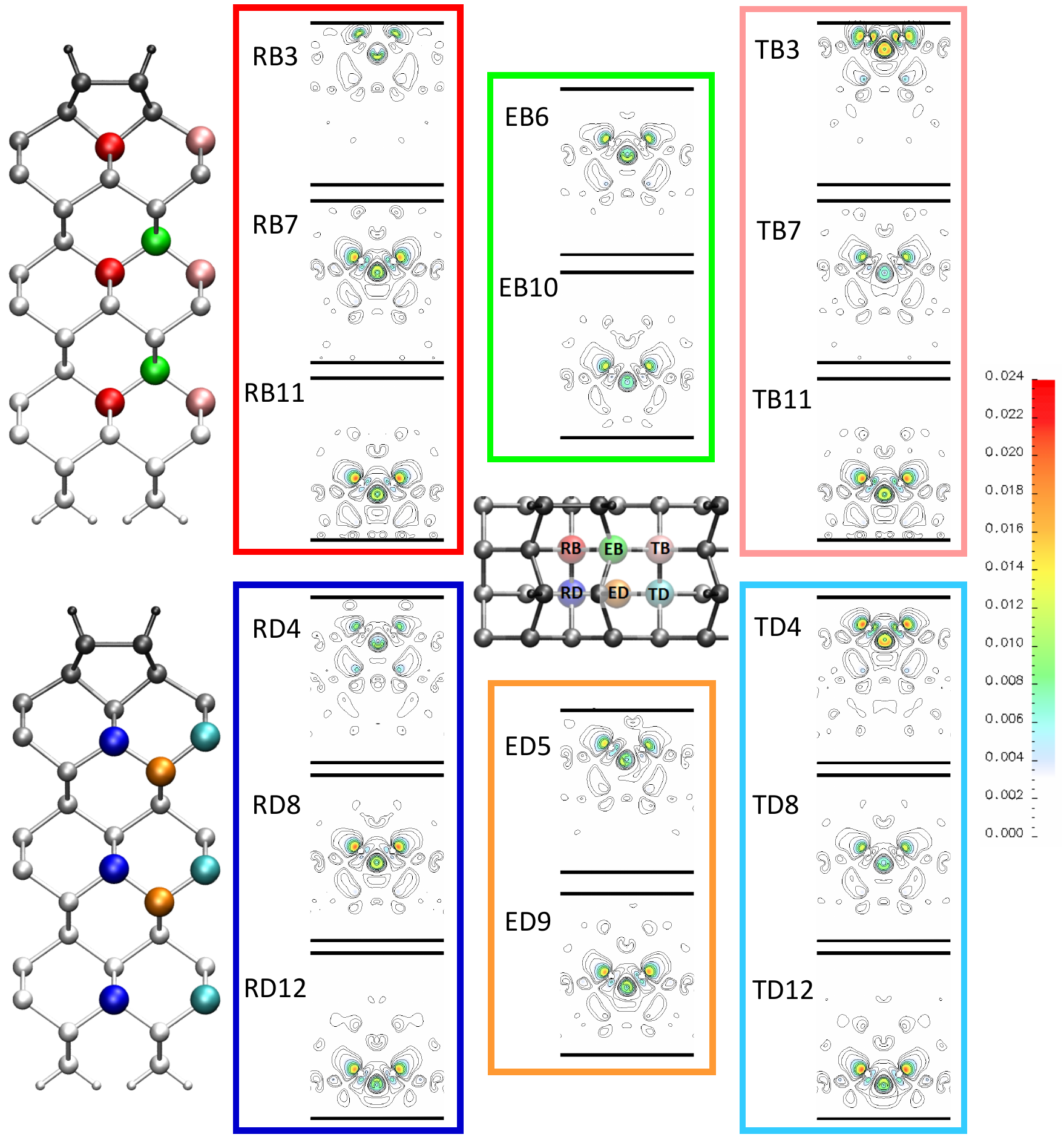}
\caption{\label{f:fig3SI}Contour plots of the charge density of the lowest impurity band, for different dopant positions. The contours are plotted on a plane perpendicular to the surface; the orientation of the plane (either along dimer rows or perpendicular to dimer rows) is chosen for each defect family so that two lobes in the charge density point towards the upper surface. A colour scale for the contour plots is given, in units of electrons per cubic angstrom.  Horizontal black lines denote the approximate top and bottom surfaces of the slab but are only a guide for the eye; for description of defect site labelling see text.}
\end{figure*}

The detailed differences within each symmetry family
can be understood when we consider the  shape of the dopant wavefunction as it interacts with the surface.  Fig.~\ref{f:fig3SI} shows contour plots of 
the charge density of the highest occupied band in the dopants (we note that the decay of the state into the vacuum is not shown as we are concerned with the effect of the proximity of the surface on the state).
All the plots share a core structure derived from the $sp^3$ hybridization of the As orbitals that point along the As--Si bonds. However, the diffuse outer parts of the states for dopants close to the surface are compressed by its proximity, with compression increasing with decreasing distance; the compression is most easily seen by comparing, for example, RB3 and RB7, where the lobes in RB7 are clearly flattened and deformed in RB3. The unique shape of each feature in STM stems from the combination of this compression, the cut taken through the state by the surface plane, and its interaction with the local atomic and electronic structure of the surface. For each symmetry family, the most distinctive feature is the one corresponding to the site nearest the surface (e.g. RB3, RD4); for deeper-lying donor sites (e.g. RB7, RB11 or RD8, RD12) the compression by the surface is less significant but there is still variation in the image that can be used to distinguish them. In the bulk the dopant families become equivalent and develop the full $T_d$ symmetry of the impurity.
(Details on the influence of additional charge on the wavefunction-surface interaction and a discussion of the supercell band-structure are found in Appendix~\ref{sec:effect-an-additional}.)

\begin{figure}
\includegraphics[width=9.0cm]{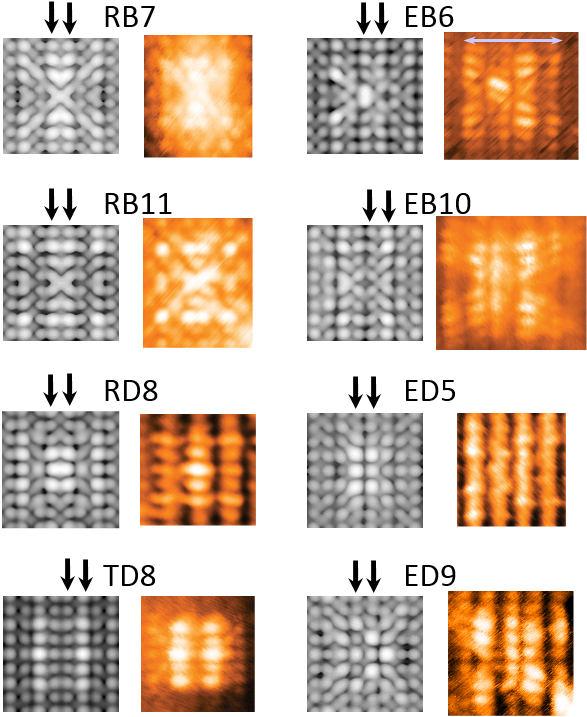}%
\caption{\label{f:fig3}(Color online) Simulated (left column, greyscale) and experimental (right column, orange) STM images of individual dopants assigned to exact sites. Pairs of black arrows indicate dimer rows. 
The pale horizontal arrow marks four dimer rows, $\sim$31~\AA. 
Simulated images were integrated from $E_F+0.086~\mathrm{eV}$ to
$E_F-0.7~\mathrm{eV}$, except for EB10, which was integrated from
$E_F+0.071~\mathrm{eV}$ to $E_F-0.7~\mathrm{eV}$. Experimental images 
were recorded at bias voltage of $-$1.3~V except for RB7 and EB10,
recorded at $-$1.2~V.}
\end{figure}

Using our DFT results we can therefore interpret the experimental images and precisely locate each dopant using the differences in images within each family (Figure~\ref{f:fig3}).
On this basis we have identified pairs of corresponding simulated and experimental images for eight distinct sites (Figure~\ref{f:fig3}) corresponding to five of the six symmetry families. The theoretical unit cell corresponds to the central part of each experimental image, and the agreement in this region is good even for the more diffuse images of the deeper dopants.
We identify the high symmetry sites (left side of Figure~\ref{f:fig3}) unambiguously; the lower symmetry sites are more challenging, but we are nevertheless able to find candidate images from layer 5 to layer 10 (right side of Figure~\ref{f:fig3}).
As discussed above, our identification rests on symmetry, size and general appearance.  The key factor is symmetry: the location of the dopant relative to the dimer row and dimers within the row.  The size and appearance are determined by the arrangement of the brightest spots within the feature and their position relative to the centre of symmetry. 

The first two sites in the left-hand column of Figure~\ref{f:fig3} belong to the RB family, as they share the distinctive cross feature centred on the dimer row; the difference between the two is seen in the strong central stripe (RB7, top) and the four outer bright dots (RB11, lower).  We can eliminate RB3 as a candidate for either of these, as they are too diffuse.

For the RD and TD families, we only found one candidate for each, which are again assigned to the appropriate family on the basis of the symmetry.  RD8 is chosen by comparison with Figure~\ref{f:fig2} on the basis of the strong central feature and the relative brightness of the outer five dots on both sides (in RD8 these alternate in brightness along the row).  TD8 also shares this alternation of brightness in the outer dots, which is not seen in TD12, and lacks the strong central feature of TD4.

The second column of Figure~\ref{f:fig3} is more challenging.  All of the experimental features clearly show the asymmetry characteristic of the EB and ED families, and the assignment to B or D is made on the basis of the position of the centre of the feature relative to the surrounding dimers.  We explain here our reasoning in assigning the layers (though these are not unambiguous).

We assigned EB6 on the basis of the two horizontal v-shapes in the centre of the image (resembling $>$ and $<$) seen clearly in the simulated image, and on its size: it is confined to relatively few dimers.  EB10 shows least asymmetry of any of these features in both experiment and modelling, as well as a central cross shape (slightly separated vertically) which matches well with simulations.

ED5 shows six asymmetric, central dots in simulations, which are seen in the STM image, and are on the same dimer row in both.  The outer dots also show a similar pattern in experiment and modelling.  The final image, assigned to ED9, shows an overall horizontal v-shape (resembling $>$) and extends over several dimers and dimer rows, and it is on this basis that the assgnment is made.

We have not observed any features corresponding to layers 1-4; this is likely because the annealing process removes dopants from these layers.\cite{a:Wolkow2012}

\begin{table}
\caption{\label{t:energies}Relative energies of the different dopants (negatively charged cell, see Methods for details). All values are in eV.}
\begin{ruledtabular}
\begin{tabular}{ll}
Site & $\Delta(E)$ \\
\hline
RB3  & 0.0\footnote{Reference energy}\\
TB3  & 0.14 \\
TD4  & 0.11 \\
RD4  & 0.04 \\
ED5  & 0.04 \\
EB6  & 0.04 \\
RB7  & 0.01 \\
TB7  & 0.07 \\
TD8  & 0.08 \\
RD8  & 0.03 \\
ED9  & 0.04 \\
EB10 & 0.07 \\
RB11 & 0.07 \\
TB11 & 0.05 \\
TD12 & 0.16 \\
RD12 & 0.16 \\
\end{tabular}
\end{ruledtabular}
\end{table}

In many systems, the relative energies of different structures can be used to determine which structure is most likely. In this work the experimental procedure to prepare the samples is expected to obscure the effect of the relative energies. We have, nevertheless, investigated the energetics of the theoretical models as well.
The total energies of all the dopant structures after relaxation are shown in Table~\ref{t:energies}.
The most stable dopant site is RB3; other sites are up to 0.20 eV less stable, depending largely on proximity to the bottom of the slab, and the artificial surface and fixed atoms associated with it.  
There are weak trends that can be discerned, in particular that in layers  near the surface where both row and trench positions are possible (layers 3, 4, 7, 8) the row position is more stable than the trench position.  Contour plots of the charge density of the dopant state (shown in Fig.~S2 in SI)  
confirm this trend: while there are differences between all sites depending on their location, the core is always very similar; the shallow row/trench sites (L3, 4, 7, 8), however, are strongly affected by the surface, and differ between row and trench, and with depth.  The influence of the surface on the dopant state is clearly significant.
The energy differences are close to the limits of practical DFT accuracy and are not expected to play an important role in determining the site populations in our sample, which had been annealed to $800$~$^\circ\mathrm{C}$ for 200~s prior to the experiment (see Methods).

\section{Conclusions\label{sec:conclusions}}
We have demonstrated that the dopant state of a neutral Group V donor is extremely sensitive to the exact position of the donor, and gives a unique appearance in STM due to its interaction with the surface.  We identified experimentally observed As dopants 
below the Si(001):H surface with atomic precision, assigning a range of defect sites between 5 and 15~\AA\/ beneath the surface H atoms, by a combination of density functional simulations and scanning tunnelling microscopy.  The characteristic signature of each feature arises from the modification of the dopant wavefunction by its interaction with the surface.  This gives us an understanding of the transition between the bulk dopant (with its delocalised hydrogenic orbital) and the previously studied dopants in the surface layer.  Since this work opens the door to atomic-scale structural interrogation of buried dopants, 
it provides a key tool for any technology that require precisely placed dopants. 
The most exciting aspect of this work is that it paves the way for probing the interactions between buried dopants, towards quantum information processing applications. Applying the capabilities demonstrated here to study pairs of buried dopants of different geometries and separations will allow us not only to learn about how the dopants interact with one another, but also to test the limits of STM and DFT for the extraction of such information. We envisage a wealth of fundamental studies into the mechanisms of tunnelling into and from dopant orbital states, and the extension of DFT to account for exchange effects, accurate modelling of Rydberg states and charge transfer.


\begin{acknowledgments}
The calculations were done at the University College London High
Performance Computing cluster (Legion) and at the UCL London Centre 
for Nanotechnology HPC service. The work was funded by the EPSRC under
grant EP/H026622/1.
\end{acknowledgments}


\appendix

\section{Bandstructures\label{sec:bandstructures}}
We present details of the band structure within the unit cell we use, to enhance understanding of the effects of periodic boundary conditions required by the simulation method.

\begin{figure*}[h]
\includegraphics[width=8.0cm]{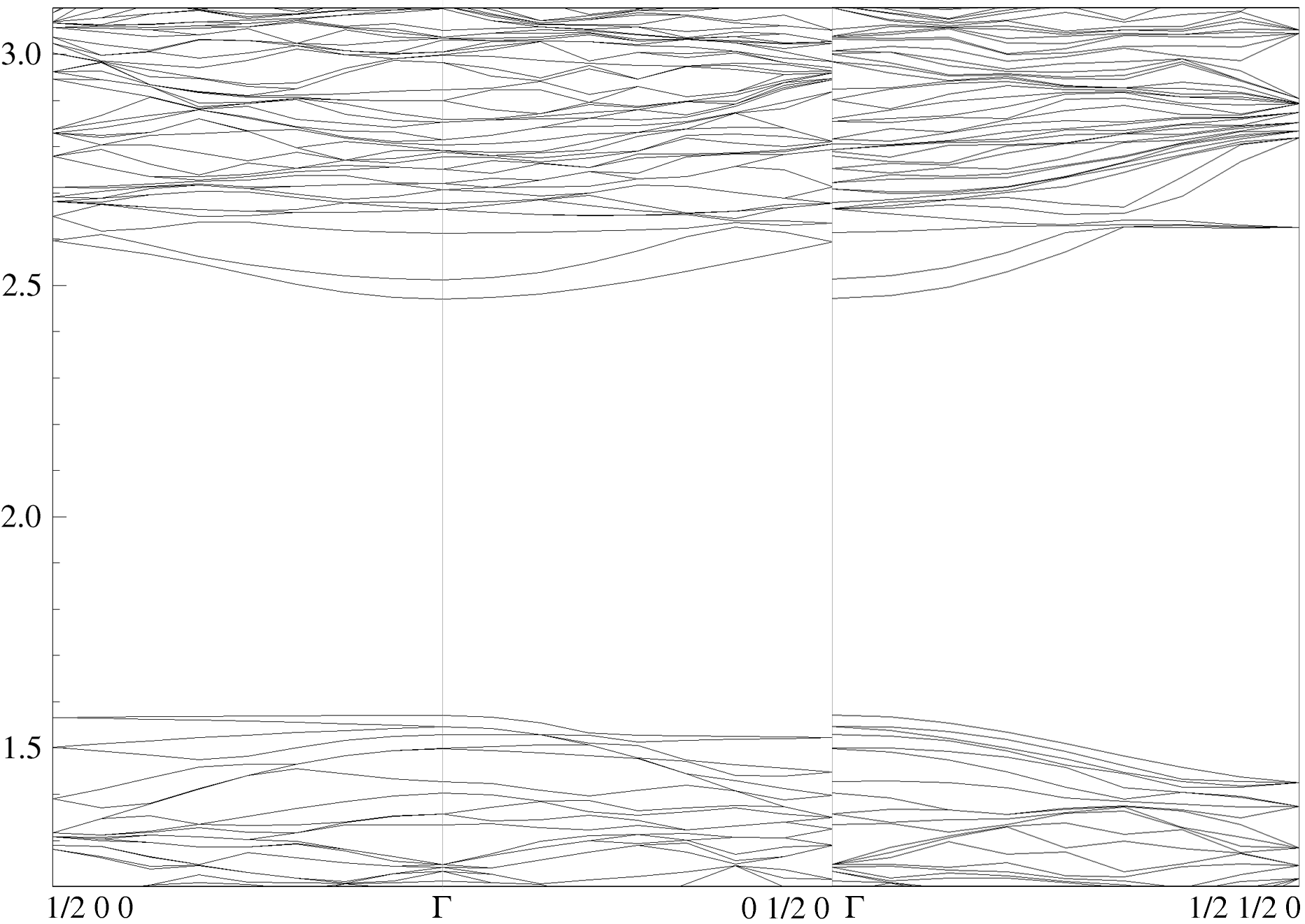}%
\hspace{0.6cm}
\includegraphics[width=8.0cm]{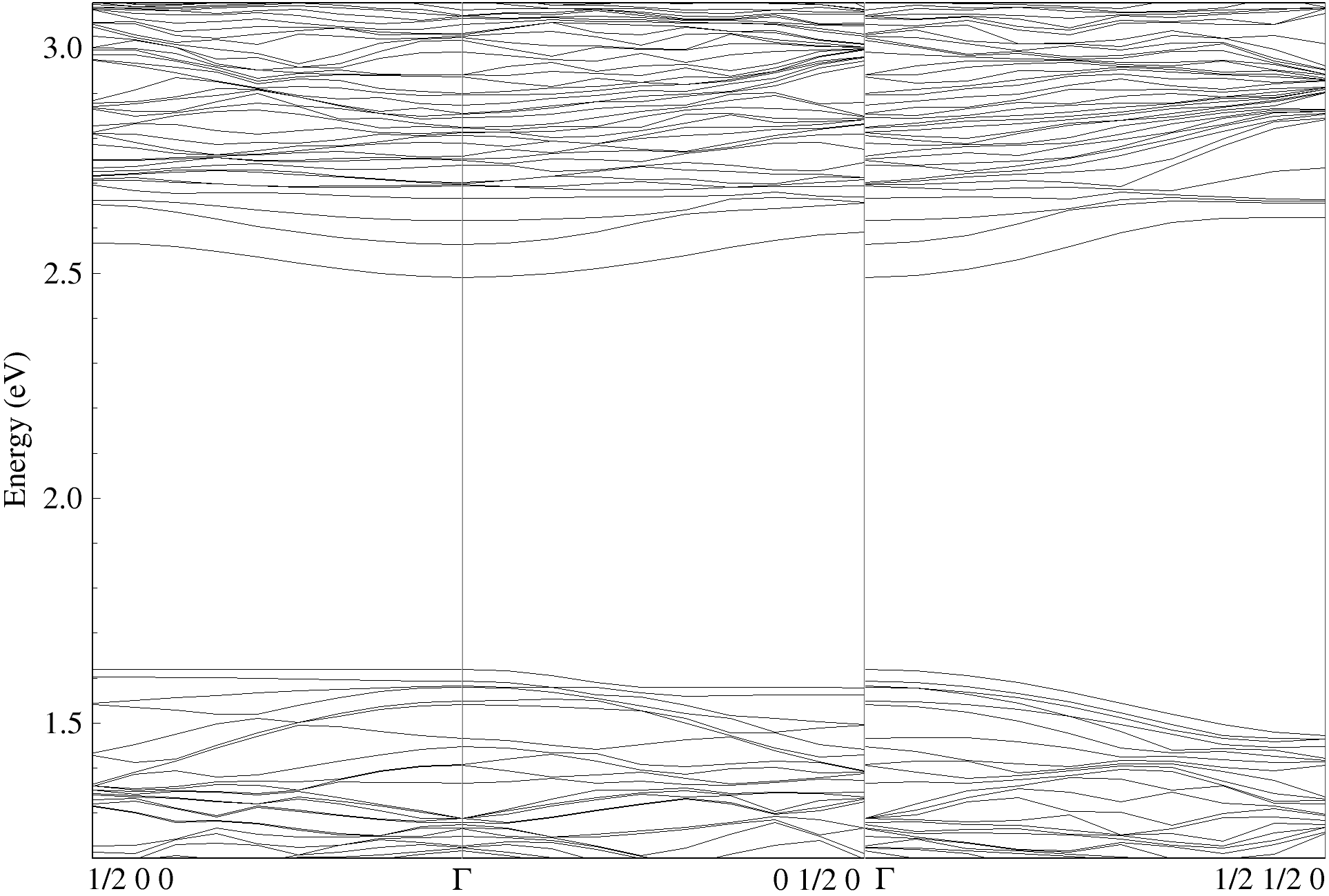}%
\caption{\label{f:fig2SI}A simulated bandstructure of the clean, neutral surface (left) and of the RB3 model (right).  The Fermi level is at 2.03~eV for the clean surface and 2.55~eV for the RB3 model. Variation of the Kohn-Sham energies is shown as a function of wavevector parallel to the surface, in units of $2\pi/A$ where $A=4\sqrt{2}a$ is the supercell in-plane lattice vector. Vectors in reciprocal space are labelled according to the reciprocal lattice vectors of the slab calculation; within the slab coordinate system, the dimer rows run along the $\langle010\rangle$ direction.}
\end{figure*}

The Kohn-Sham bandstructures for the clean surface and for a representative defect geometry (the RB3 site) are shown in Figure~\ref{f:fig2SI}. The conduction-band minima in the bulk lie between the $\Gamma$ and X points at approximately $0.85(\pm 1,0,0)2\pi/a$, $0.85(0,\pm 1,0)2\pi/a$ and $(0,0,m\pm 0.85)2\pi/a$ where $m$ is an integer; the slab boundary conditions effectively force the Kohn-Sham states to vanish on two planes with a separation of approximately $3.75a$, picking out bulk states with $k_\perp\approx n\pi/(3.75a)\,(n=1,2\ldots)$.  The lowest-energy states consistent with these boundary conditions are derived from the $(0,0,m\pm 0.85)2\pi/a$ bulk minima and hence have $k_\parallel=0$; two such states are evident in Figure~\ref{f:fig2SI}(a).

The potential arising from the donor causes impurity bands to form from linear combinations of the low-lying conduction band states, pulling them down into the band gap.  In the limit of infinite defect separation the width of these bands would drop to zero, giving a clear separation between localised bound states and scattering states; for the finite supercells accessible to our calculations the impurity bands retain a finite width and the distinction is not clear-cut.  Nevertheless a single band can be identified with its minimum at the $\Gamma$-point, which lies below all the other states.  The band dispersion (approximately 0.1~eV) is large compared with the expected spacing of the Rydberg series associated with the hydrogenic impurity (30~meV or less) so only the lowest state is clearly resolved below the continuum; nevertheless this is the state most relevant to low-temperature measurements.

For the clean surface the conduction-band states are all empty and the Fermi energy lies in at mid-gap.   For the defect system we used a negatively charged cell to mimic the surface in the presence of tip-induced band bending to simulate the STM images, as explained in Methods; a total of two electrons must therefore be accommodated in the impurity states.  As a result the Fermi level lies just above the bottom of the conduction band, at 2.55~eV on the scale of Figure~\ref{f:fig2SI}(b).

 \section{Effect of Fermi level and $k$-point sampling\label{sec:effect-fermi-level}}
\begin{figure}[hbt]
\includegraphics[width=8.5cm]{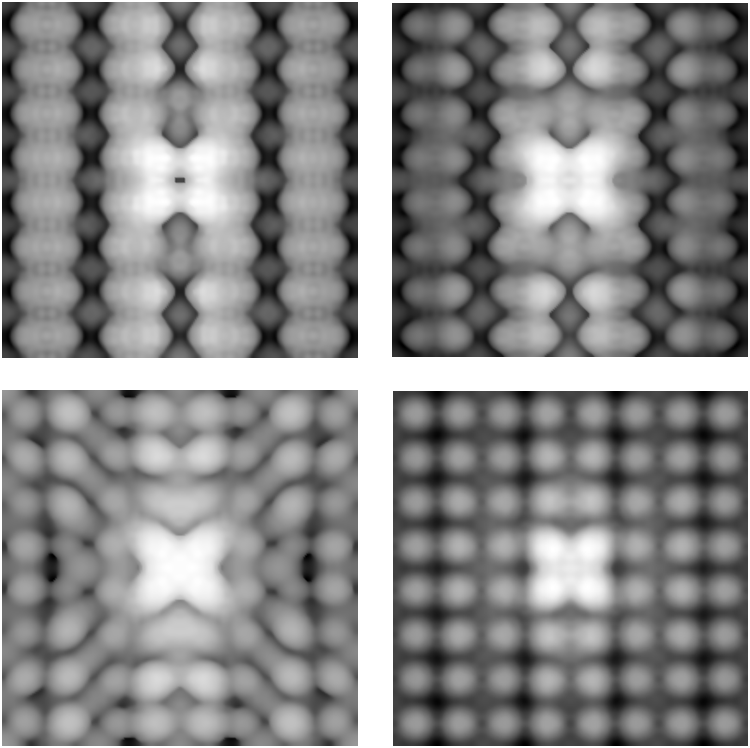}%
\caption{\label{f:fig1SI}A comparison of simulated STM images of the RB3 defect created with different parameters. Top left: $K$-point mesh is shifted to include the $\Gamma$ point, integration window $E_F$ to $E_F-0.7~\mathrm{eV}$, thermal smearing 0.1~eV. Top right: K-point mesh shifted to include the $\Gamma$ point, integration window $E_F$ to $E_F-0.7~\mathrm{eV}$, thermal smearing 0.01~eV. Bottom left: K-point mesh does not include the $\Gamma$ point, integration window  $E_F+0.086~\mathrm{eV}$ to $E_F-0.7~\mathrm{eV}$, thermal smearing 0.1~eV. Bottom right: Neutral system. K-point mesh shifted to include the $\Gamma$ point, integration window $E_F$ to $E_F-0.7~\mathrm{eV}$, thermal smearing 0.1~eV.}
\end{figure}

Fig.~\ref{f:fig1SI} shows the influence of the detailed computational setup on the simulated STM images. Experimental filled-state images are obtained with negative sample bias; hence in simulated images, electrons should tunnel from a window of states just below the Fermi energy (from $E_F$ to $E_F-\mathrm{e}V_\textrm{bias}$ at low temperature, where $E_F$ is the computational Fermi energy).  The $(2\times 2\times 1)$ Monkhorst-Pack $k$-point grid used elsewhere in this paper does not include the $\Gamma$-point and the lowest states therefore lie just above $E_F$; hence an STM integration window with an upper limit set to the Fermi energy $E_F$ does not contain any bands at the sampled $k$-points and does not produce a useful STM image. We have therefore shifted the integration window, typically so its upper limit is at $E_F+0.086~\mathrm{eV}$ (Fig.~\ref{f:fig1SI} bottom left). Alternatively we can change the $k$-point sampling so that the $\Gamma$ point is included; bands are then present in an integration window from $E_F$ down to $E_F-0.7~\mathrm{eV}$ (Fig.~\ref{f:fig1SI} top left) and a similar image is produced. If we also decrease the Gaussian smearing of occupations to 0.01~eV, the Fermi level moves up to maintain the total electron population; so we include more bands (Fig.~\ref{f:fig1SI} top right) when integrating from the Fermi level, and the resulting image begins to display the characteristic central $\times$ typical of images obtained with a shifted integration window (Fig.~\ref{f:fig1SI} bottom left). On the other hand, in a neutral system with a single electron in the impurity band (Fig.~\ref{f:fig1SI} bottom right) the bands that give rise to the STM feature once again shift above the Fermi level.

Because of the substantial band-bending and the partial screening of the tip field by mobile charges, and DFT errors in the Kohn-Sham eigenvalues of unoccupied states, the precise relationship between the applied voltage and the theoretical voltage bias window is complex.  In practice it is determined semi-empirically by the need to include the most important defect-related states.   The parameters used to generate the STM images in the paper were as follows.  For Figure~1 in the main text: 
simulated images were integrated  from $E_F+0.086~\mathrm{eV}$ to $E_F-0.7~\mathrm{eV}$, and experimental images were recorded at sample bias voltages of $-$1.2~V (Fig.~1a) and $-$1.3~eV(Fig.~1b).  For Figure~2 in the main text: simulated images were integrated from 
from $E_F+0.086~\mathrm{eV}$ to $E_F-0.7~\mathrm{eV}$, except for the following: EB10: $E_F+0.071~\mathrm{eV}$ to $E_F-0.7~\mathrm{eV}$, TD12: $E_F+0.065~\mathrm{eV}$ to $E_F-0.7~\mathrm{eV}$, and RD12: $E_F+0.063~\mathrm{eV}$ to $E_F-0.7~\mathrm{eV}$.  For Figure~3 in the main text: 
simulated images were integrated from $E_F+0.086~\mathrm{eV}$ to $E_F-0.7~\mathrm{eV}$, except for EB10, which was integrated from $E_F+0.071~\mathrm{eV}$ to $E_F-0.7~\mathrm{eV}$.  Experimental images were recorded at the following bias voltages: RB7: $-$1.2~V, RB11: $-$1.3~V, RD8: $-$1.3~V, TD8: $-$1.3~V, EB6: $-$1.3~V, EB10: $-$1.2~V, ED5: $-$1.3~V, ED9: $-$1.3~V.

\section{Effect of an additional electron on the surface-wavefunction iteraction\label{sec:effect-an-additional}}
\begin{figure*}
\includegraphics[width=15cm]{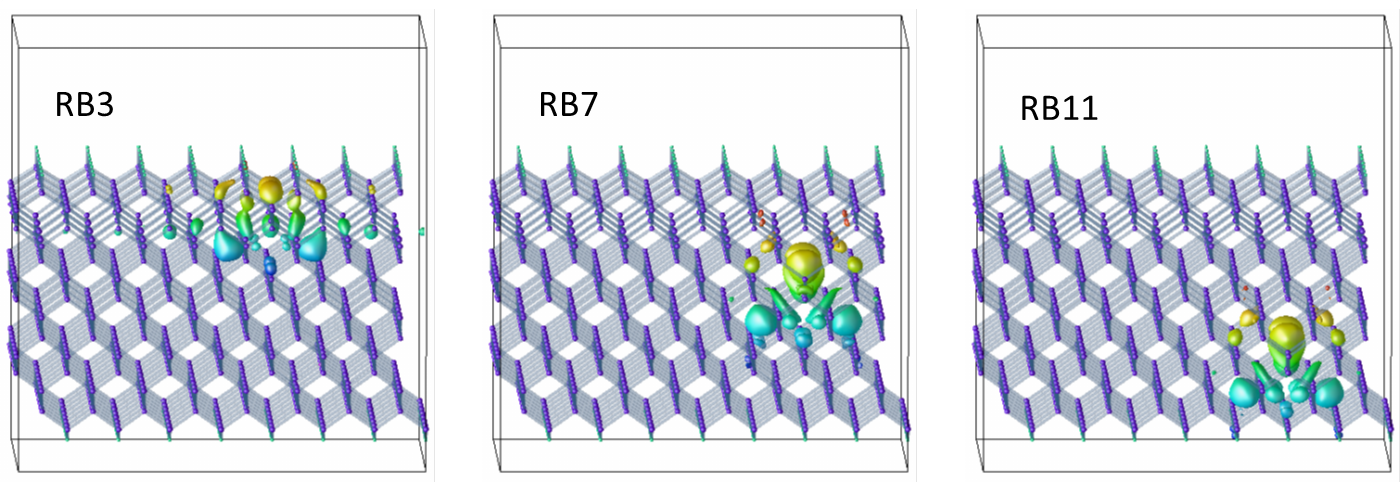}%
\caption{\label{f:fig4SI}Charge density differences between the neutral doped and undoped systems (shown at isosurface density of 0.001~eV) for the RB family of features. The extra electron had not been added in this case, so as to obtain a system with unpaired spin.}
\end{figure*}

The trends in the wavefunction interaction with the surface, described in the
main text, can also be observed in the total electronic charge difference between the neutral doped and undoped systems (see Figure~\ref{f:fig4SI}), demonstrating that the addition of the extra electron (representing the screening charge) does not change the essential physics.

\end{document}